**Title:** Effect of Toxic Review Content on Overall Product Sentiment

**Author Name and Affiliations:**

**Mayukh Mukhopadhyay**

(Corresponding Author)

Executive MBA, Vinod Gupta School of Management, IIT Kharagpur

ORCID: 0000-0001-7733-7160

**Institute Email:** mayukhmukhopadhyay@iitkgp.ac.in

**Sangeeta Sahney**

Professor, Vinod Gupta School of Management, IIT Kharagpur

**Institute Email:** sahney@vgsom.iitkgp.ac.in

**CRediT authorship contribution statement:**

Mayukh Mukhopadhyay: Conceptualization, Methodology, Investigation, Writing

Sangeeta Sahney: Supervision.

**Conflict of Interest Statement:** The authors declare no conflict of interest.

**Funding Statement:** This research received no specific grant from any funding agency in the public, commercial, or not-for-profit sectors.

**Acknowledgement:**

The authors would like to thank Google Jigsaw team for providing researcher access to perspective API. The authors are indebted to IIT Kharagpur Central Library for providing remote access to Scopus database for carrying necessary literature survey in midst of pandemic lockdown.

# Effect of Toxic Review Content on Overall Product Sentiment

## Abstract

Toxic contents in online product review are a common phenomenon. A content is perceived to be toxic when it is rude, disrespectful, or unreasonable and make individuals leave the discussion. Machine learning algorithms helps the sell side community to identify such toxic patterns and eventually moderate such inputs. Yet, the extant literature provides fewer information about the sentiment of a prospective consumer on the perception of a product after being exposed to such toxic review content. In this study, we collect a balanced data set of review comments from 18 different players segregated into three different sectors from google play-store. Then we calculate the sentence-level sentiment and toxicity score of individual review content. Finally, we use structural equation modelling to quantitatively study the influence of toxic content on overall product sentiment. We observe that comment toxicity negatively influences overall product sentiment but do not exhibit a mediating effect over reviewer score to influence sector-wise relative rating.

## 1. Introduction

Trolling and cyberbullying is a community driven online behaviour that effects specific individuals participating in that community (Cheng et. al., 2017). With the advent of world wide web and social network sites, trolls have taken the form of online toxic comment. A comment is perceived to be toxic when it is rude, disrespectful, or unreasonable and make individuals leave the discussion (Noever, 2018). Various researchers have conducted extensive study to identify such patterns of online trolling.

## 2. Bibliometric Analysis

A concise bibliometric study was done to identify impactful publications related to online troll and comment toxicity (OTTC). The query was made between the time period of 08-Aug-2020 to 18-Aug-2020. The motive of having a 10-day window was to check whether the corpus count varied with any new literature addition. Figure 1 depicts the Scopus database search query used along with the document count of OTTC corpuses. SCOPUS database was chosen for the following three reasons:

- The batch size of metadata retrieval is 2000 compared to Web of Science whose batch size is 500 articles.
- The multidisciplinary base of SCOPUS is wider than public catalogues like arxiv which has subject specific preprints.
- SCOPUS primary collection comprise of peer-reviewed publications when compared with google scholar which also has literature indexed from predatory journals or unreviewed preprints.

| SL NO | Corpus Title | Search Query | Document Count |
|---|---|---|---|
| 1 | OnlineTroll_ToxicComment [OTTC] | TITLE-ABS-KEY ( "Perspective API" OR "Toxic Troll" OR "Toxic Comment" OR "Online Trolling" OR "Cyber Bullying" OR "Online Harassment" OR "Online Bullying" OR "Cyber Harassment" ) | 1241 |

Figure 1. Scopus Database Search Query

The operator TITLE-ABS-KEY searches the string in quotes from the title, abstract and author's keyword from the Scopus listed documents to return the most relevant output with decreasing order of citation count. We retrieved a total of 1241 documents for OTTC corpus. The documents were extracted in Bibtex format that is compatible with R data ingestion. Figure 2 depicts the various parameters which were selected while document retrieval. The orange dot signifies the selected parameter while the white signifies the ignored parameter before bibtex file export from Scopus database.

| Citation information | Bibliographical information | Abstract & keywords | Funding details | Other information |
| --- | --- | --- | --- | --- |
| ■ Author(s) | ■ Affiliations | ■ Abstract | ☐ Number | ☐ Tradenames & manufacturers |
| ■ Author(s) ID | ■ Serial identifiers (e.g. ISSN) | ■ Author keywords | ☐ Acronym | ☐ Accession numbers & chemicals |
| ■ Document title | ■ PubMed ID | ■ Index keywords | ☐ Sponsor | ☐ Conference information |
| ■ Year | ■ Publisher | | ☐ Funding text | ■ **Include references** |
| ■ EID | ■ Editor(s) | | | |
| ■ Source title | ■ Language of original document | | | |
| ■ volume, issue, pages | ■ Correspondence address | | | |
| ■ Citation count | ■ Abbreviated source title | | | |
| ■ Source & document type | | | | |
| ■ Publication Stage | | | | |
| ■ DOI | | | | |
| ■ Access Type | | | | |

Figure 2. Scopus Export Parameters

After extraction of the bibtex file from Scopus database, the file was uploaded to RStudio as data frame for bibliometric analysis. R uses bibliometrix and Shiny package to visualize the data frame of bibtex file for visualization as a web-app. Figure 3 summaries the aggregated information of the bibtex file for OTTC corpus. The corpus metadata has been divided into five sections namely timeline information, document type, document content, author information and author's collaboration data.

| Description | Results |
|---|---|
| **Corpus 01** | **OTTC** |
| Timespan | 2000:2020 |
| Sources (Journals, Books, etc) | 689 |
| Documents | 1241 |
| Average years from publication | 3.72 |
| Average citations per documents | 18.08 |
| Average citations per year per doc | 2.703 |
| References | 47670 |
| **DOCUMENT TYPES** | |
| article | 838 |
| article in press | 3 |
| business article | 1 |
| conference paper | 380 |
| data paper | 1 |
| editorial | 6 |
| erratum | 1 |
| note | 10 |
| short survey | 1 |
| **DOCUMENT CONTENTS** | |
| Keywords Plus (ID) | 3774 |
| Author's Keywords (DE) | 2669 |
| **AUTHORS** | |
| Authors | 3070 |
| Author Appearances | 3953 |
| Authors of single-authored documents | 182 |
| Authors of multi-authored documents | 2888 |
| **AUTHORS COLLABORATION** | |
| Single-authored documents | 196 |
| Documents per Author | 0.404 |
| Authors per Document | 2.47 |
| Co-Authors per Documents | 3.19 |
| Collaboration Index | 2.76 |

Figure 3. OTTC Corpus Aggregated information

For corpus OTTC we observe in figure 4 that the core literature document mainly comprises of articles and conference papers with a compound annual growth rate (CAGR) of 30.59% which is consistent with our motivation to work on effects of toxicity scoring.

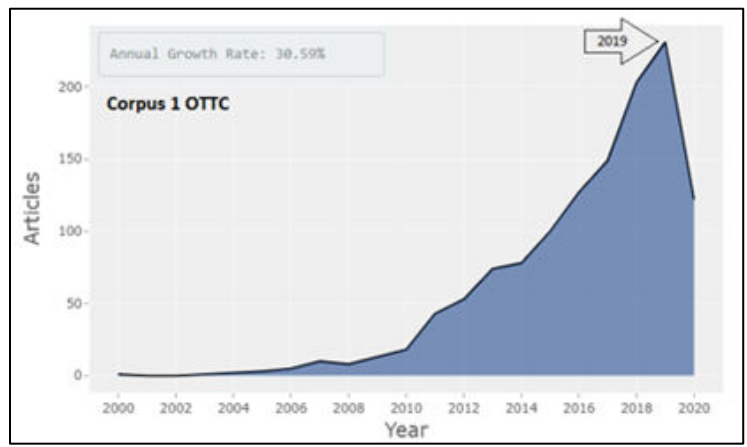

Figure 4. OTTC Corpus Compounded Annual Growth Rate

To identify the most impactful contributions in the OTTC corpus, we focused on five types of analysis - source clustering using Bradford's law (Bradford, 1976) as depicted in figure 5, single country and multi-country publications in figure 6, author productivity timeline in figure 7 , country-author-keyword Sankey plot in figure 8, and author collaboration network in figure 9.

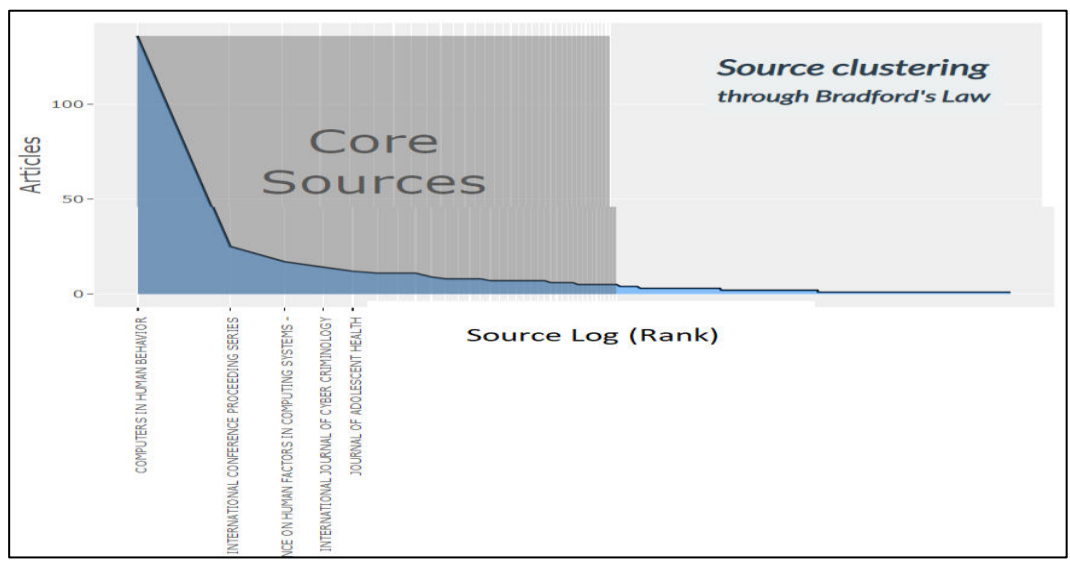

Figure 5: Source cluster of Top 5 journal of OTTC corpus

We observe that top 5 journals contributing to OTTC literature are Computers in Human Behaviour (CHB), International Conference Proceeding Series, Conference on Human factors in computing system, International Journal of Cyber Criminology and Journal of Adolescent health.

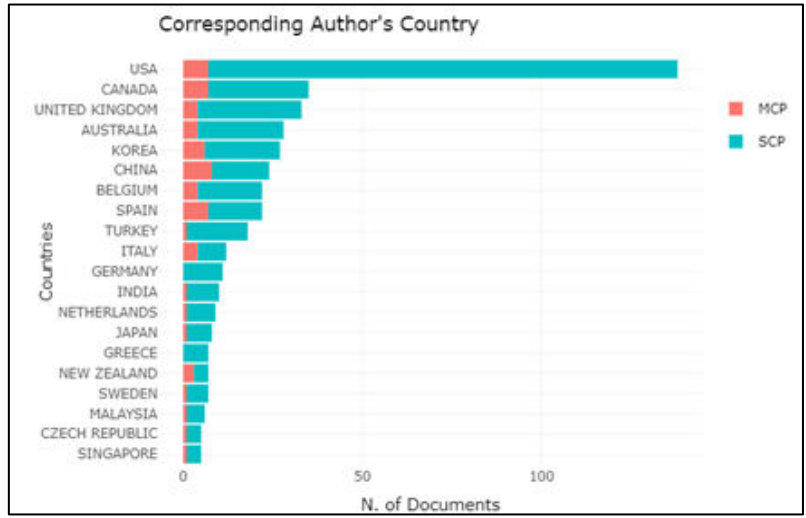

Figure 6. Single-Country and Multiple-Country Publications of OTTC corpus

Single country publication (SCP) denotes those work that are published by authors of same country. Multi country publication (MCP) denotes those work which involve at least one author from a different country of origin. SCP to MCP ratio directly correlates to the collaborative index of the author.

Figure 7 has three dimensions about an Author's work: Time, Colour and Size. The line denotes the timeline, the intensity of colour represents the impact of the work in terms of citation and the size of the blobs on the timeline represent the percentage of work done on the timeline.

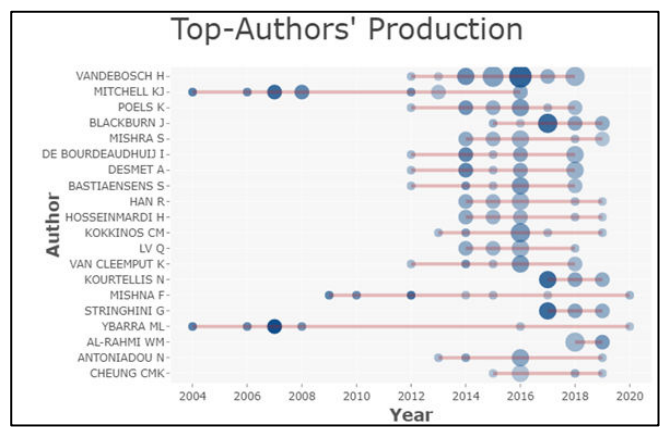

Figure 7: Top 20 Authors' Production Timeline of OTTC Corpus

The Sankey plot is a three-way-plot that links the country, author, and keyword in the OTTC corpus. Belgium stands out on most of the impactful work done in OTTC corpus, specializing in cyberbullying phenomenon.

It is to be noted that as part of document content we have a new parameter called Keyword Plus along with Author's Keyword. Keyword Plus is a Scopus database feature where an inbuilt machine learning algorithm scan the references of a document and generates keywords that are not present in the keyword specified by Author or absent in the title and abstract. Garfield (1993) claimed that Keyword Plus terms can capture an article's content with greater depth and variety. Zhang et al. (2016) countered that although keyword plus is as effective as Author's keyword in terms of bibliometric analysis investigating the knowledge structure of corpus fields, but it is less comprehensive in representing an article's actual content.

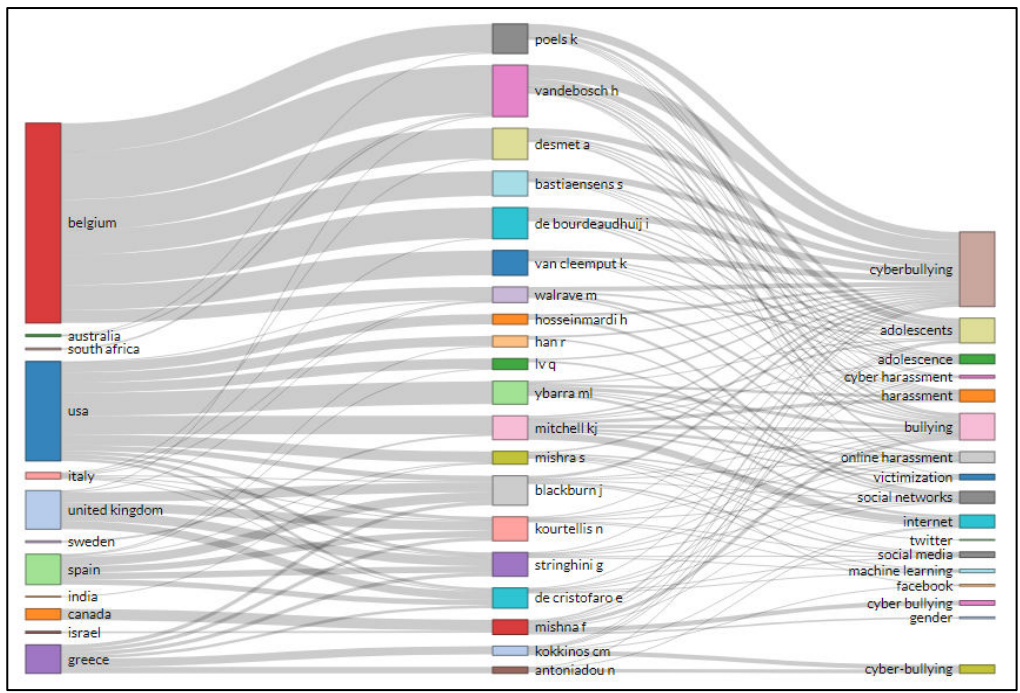

Figure 8. Country-Author-Keyword Sankey Plot of OTTC Corpus

For discovering knowledge cluster, social collaborative network approach has been visualized in figure 9. This approach was proposed by Batagelj & Cerinsek (2013) and Aria & Cuccurullo (2017). In our visualization, independent nodes has been suppressed for highlighting collaborative work.

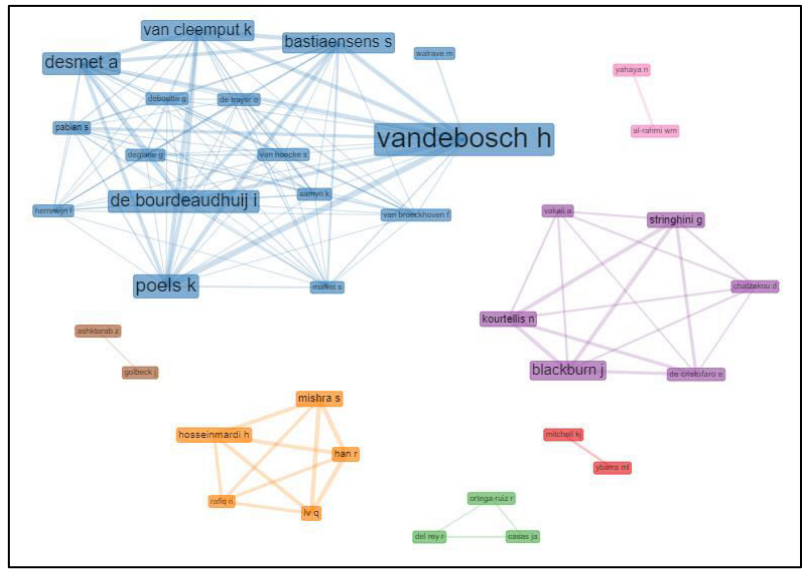

Figure 9. Author's Collaboration Network of OTTC Corpus

Based on the knowledge cluster visualization, most impactful research was done by Bastiaensens et al (2014) who underwent a contextual experimental study of bystander behaviour that intent to help victims or reinforced the online bullying. Studied over 453 secondary school students, their work revealed that bystanders had high intents to join the online bullying when other bystanders were not just acquaintances but good friends. However, the intention to help the bullied victim were witnessed during severe incidents. Furthermore, gender of the victim influenced both reinforcing and opposing behavioural intention of the bystanders. Hosseinmardi et al (2015) did extensive studies on labelling cyberbullying data using Instagram and Ask.fm social network where they crowdsourced the comment labelling activity into host of features like profanity, cyberaggression, and comments that lead to suicidal behaviour among American teens. Desmet et al (2016) showed that school or class factor were not strong predictors of bystander behaviour during cyberbullying in adolescents. The study also revealed key insights on the multifaceted nature of positive and negative intent to online bullying. The facets of negative intent attributed to passive by-standing, joining, and reinforcing due to lack of social, empathetic, and coping skills while the facets of positive intent attributed to defending, comforting, and reporting. Vranjes et al (2017) extended their study towards workplace cyberbullying that attributed work-stress events leading to such intent. They proposed discrete emotions of anger, sadness and fear played important role in explaining such behaviour which aligned with the Affective Events Theory (Weiss & Cropanzano, 1996). Chatzakou et al (2017) provided a principled and scalable approach to detect aggressive and bullying behaviour on twitter microblogging platform. using 1.6 Million twitter datasets, they showed that classification algorithm powered by machine learning statistical principles can detect aggression and bullying behaviour with 90% accuracy.

Adding to the extant literature, our study is the first attempt to develop a structural equation model using alternative user generated data to study the interaction of comment toxicity on overall product sentiment.

The article has been divided into three part. In first part we discuss various theoretical frameworks that led us to form our research questions. In the second part we discuss our methodology of data extraction, enrichment, and model evaluation. In the final section we revisit to answer our research questions using the interpretations of the evaluated model and conclude with the practical implication, contribution, limitation, and future scope of our study in understanding consumer psychology in the marketing domain.

## 3. Background and research questions

### 3.1 Alternative Data

With the boom in data literacy and advancement in data collection technologies, we have seen a booming sector of tapping alterative data to derive insights. Alternative data is basically data derived from external sources, essentially anything outside that a company typically releases.

Alternative data has three fundamental characteristics: unexplored, unrecognized, and non-traditional. By unexplored it means there is a lack of standardized metric to measure the usability of the data. By unrecognized it means such data are usually by-product of some other activity that is monitored in a standard fashion. By non-traditional it means that it is hard to process such data using traditional data processing tools and often requires specialized algorithms. Broad examples of alternate data are Internet-of-Things (IoT) sensor log data, Natural Language Processing (NLP) web data, and Environment-Societal-Governance (ESG) data used for financial analysis and reporting.

Alternative data like online reviews act as a vital source of information for consumers, especially the ones who are uninformed (Chatterjee, 2019). While considering alterative data for our present study we have focussed on mainly three parameters: to know who our data provider is working with (exclusivity), data is used as proxy indicator or false assumption (correlation), and expert tools available to interpret the data from alternative sources (application). Most of the use-cases of alternative data is based on relevance to the moment.

## 3.2 Comment Toxicity

Since many scientists lamented the scarcity of published literature on toxic content (Djuric, et al. 2015), subsequent researchers have sought to identify abusive text patterns and potentially block online trolls, partly because they result in real harm but also because they curtail further online spread.

Pavlopoulos, et al. (2017) diagnosed statistical models with word-vector embeddings on different online Greek and English commentaries on Wikipedia. Parekh, et al. (2017) reviewed many statistical models for identifying toxic commentary pattern and noted them in a binary classification of "toxic versus non-toxic". Those models were mainly support vector machine, Naïve Bayes classifier, and linear regression.

In one recent collaboration, Jigsaw's new deep learning model, called Perspective application programming interface (API), provides a comment moderation tool to detect toxic behaviour. The algorithm formulates its toxicity scores using a convolutional neural net (CNN). To improve the scoring, Wikipedia, Jigsaw, and Google released a set of user generated content to international data community via Kaggle (Noever, 2018).

David Noever (2018) further evaluated sixty-two statistical models representing nineteen major algorithmic families against attributes extracted from the Kaggle dataset and compared the models based on statistically significant differences in relative execution time and accuracy.

Jigsaw Counter-Abuse Technology Team compiled the Toxic Comment Classification data by capturing user generated comments from Wikipedia Discussion Pages as training dataset alongside a similar but unlabelled test dataset. As rated by human Turks, the trained labels included seven kinds of comment with varied annotations.

However, the research community provided deeper criticism of bias in machine learning models (Greenberg, 2017) used in perspective API. The community attributed such an outcome by potentially proving deeply entrenched bias when human Turks label and censor offensive content. Some criticism stemmed rightly from mimicked patterns of gender, ableism and racial bias found in toxic labelled content while a core research debate followed from the model's pseudo-transparent reasoning itself.

**3.3 Service Relative Rating**

Improving customer satisfaction requires a different perspective than doing a market survey. There are three broad types of collecting customer feedback namely feedback form, review request, and market research form.

With regards to the intangibleness of tourism products, an online review in the travel sector has been given more importance than a review of subscription based general product (Shin et. al.,2016). Near-future tourists are more inclined towards concrete reviews while far-future tourists are influenced more by abstract reviews (Chatterjee, 2019). Even pharmaceutical sector has a higher importance than general products due to involvement of third parties like doctor's prescription and pharmacist perception (Liekens et. al., 2012) but lower than the travel sector. Based on these heuristics, we have anchored that consumers have to travel small psychological distance while reviewing general products in the subscription sector, large traversal in travel sector, and an adjusted medium traversal in the pharma sector.

Trope and Libermann (2010) had provided a unified theory of a person's reaction to cognitive objects and subsequent decision making mechanism using a social cognitive theory called construal level theory (CLT) in terms of psychological distance (PD).Although many PD measuring scales are present in the field in social psychology, they have been mostly limited to intrapersonal, interpersonal, and organizational level (Chen & Li, 2018). In our study, we heuristically assume a scale of low (1 - service), medium (2 - pharma), and high (3 - travel) due to lack of measurement scale of PD between an individual and an industrial sector. By "low or 1" we mean that a consumer reviewing an online review content has to traverse a small psychological distance from an abstract construal level to concrete construal level for a subscription-based-service application.

In our heuristical PD, the distance travelled for service-app is the least, for pharma-app it is in middle, and for travel-app the PD travelled is the most. Hence the service cluster in our data is not a nominal exogenous variable but an ordinal exogenous variable.

As part of our study, we have also developed a sector-wise weighted rating named serviceRelativeRating (SRR) that determines the contribution of a service player towards the overall customer perception in a specific sector. In figure 10, Rp denotes overallProductRating of a player in google play-store and Rc denotes reviewerCredit value. For more than a million reviewer the Rc value is taken as 2 while for less than a million reviewer the value is 1.

$$sRR_i = \frac{(R_p R_c)_i}{\frac{1}{10}\sum_{i=1}^{n}(R_p R_c)_i}, [i \in \mathbb{N}]$$

Figure 10: Sector-wise weighted rating named serviceRelativeRating

The boxplot in figure 11 depicts the variation of SRR in a sector-wise manner. We observe the least spread of rating in the pharma sector. As the data is a weighted mean we have no outliers. The reviewerCredit plays a significant role in the amount of spread. We can infer that due to absence of any big player in the pharma sector w.r.t mobile application, customer perception is almost similar to various players in the pharma sector.

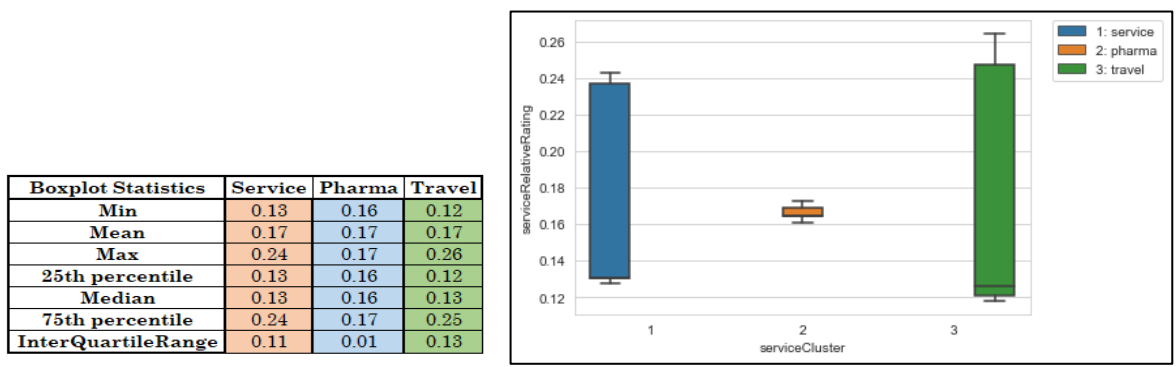

| Boxplot Statistics | Service | Pharma | Travel |
|---|---|---|---|
| Min | 0.13 | 0.16 | 0.12 |
| Mean | 0.17 | 0.17 | 0.17 |
| Max | 0.24 | 0.17 | 0.26 |
| 25th percentile | 0.13 | 0.16 | 0.12 |
| Median | 0.13 | 0.16 | 0.13 |
| 75th percentile | 0.24 | 0.17 | 0.25 |
| InterQuartileRange | 0.11 | 0.01 | 0.13 |

Figure 11: Boxplot of serviceRelativeRating

### 3.4 Research Questions

According to accessibility-diagnosticity (AD) model, consumer behaviour is driven by accessibility and diagnosticity of input variables (Chatterjee, 2019). Reviewer ratings are accessible and diagnosable which can be used as primary drivers that influence overall product sentiment. Moreover, the multiple pathway anchoring and adjustment (MPAA) model suggests that reviewer sentiments serves as a mediator in which a consumer can express their attitude towards a service taken by giving a product rating which eventually affect the sector specific rating (Chatterjee, 2019).

Psychological distance (PD) plays an important role in sector-wise rating scales to determine the construal level of a consumer while availing an app service of a particular sector (Trope & Liberman, 2010). For example, a typical consumer would frequently use a subscription-based service app on daily basis compared to leisure use of travel app or an emergency use of pharma app. In our study we provide insights into the following research questions based on AD, MPAA, PD model of consumer behaviour.

*RQ1: Does the reviewer's score influences review content sentiment and toxicity?*

*RQ2: For a given review sentiment, does review toxicity influences overall product sentiment?*

*RQ3: Does comment toxicity acts as a significant mediator between reviewer's score and sector-wise relative rating?*

*RQ4: Does overall product sentiment gets influenced by the sector of study?*

## 4. Methodology

The data workflow has been broadly divided into three phases: Extraction, Enrichment and Evaluation. In the extraction phase, a web scrapping module is developed using python to collect app review comments and related attributes from google play-store. In the enrichment phase we utilize Perspective application programming interface (API) and setup data-frame in R to calculate toxicity score of each comment. Perspective is a request-only service provided by Google's Jigsaw team to research on abusive content in user comments. Parallelly, we use sentimentr package in R to determine the aggregated sentence-level sentiment score for each review comment. In the Evaluation Phase, we append these scores with service relative rating pass it through the lavaan package in R to develop path analysis model, measurement model, total effect model and full SEM model to study the influence of comment toxicity on overall product sentiment. The project steps have been summarized in six steps:

Step1: Creation of Customer Review Scrapping tool using Python

Step 2: Scrapping google play-store app reviews from three different sectors

Step 3: For calculation of toxicity score, utilizing Perspective API interface with R.

Step 4: Calculation of individual toxicity scores of the scrapped reviews

Step 5: Calculating the sentence level sentiment score and aggregate for each review.

Step 6: Using structural equation modelling to study the interaction of the latent indicator of product sentiment and how it is affected by comment toxicity scores.

**4.1 Extraction Phase**

In the data extraction phase of our study, we have focussed on web data and NLP based information processing. We have identified three broad sectors namely subscription-based-services, pharma-and-digital-health, and travel as listed in figure 12.

| Service | Pharma | Travel |
|---|---|---|
| AmazonIndia | Netmeds | makemytrip |
| BigBasket | medlife | oyo rooms |
| WhiteHatJr | sastasundar | goibibo |
| Udemy | practo | redbus |
| Zomato | apollo | Uber |
| Groww | 1mg | booking |

Figure 12. Service, Pharma, and Travel Sector Players

Then we scrapped app-review data from google play-store from six well-known players from each sector (figure 13). To ensure a balanced dataset, we have collected 1200 review comments from each player i.e. 1200 x 18 = 21600 review comments.

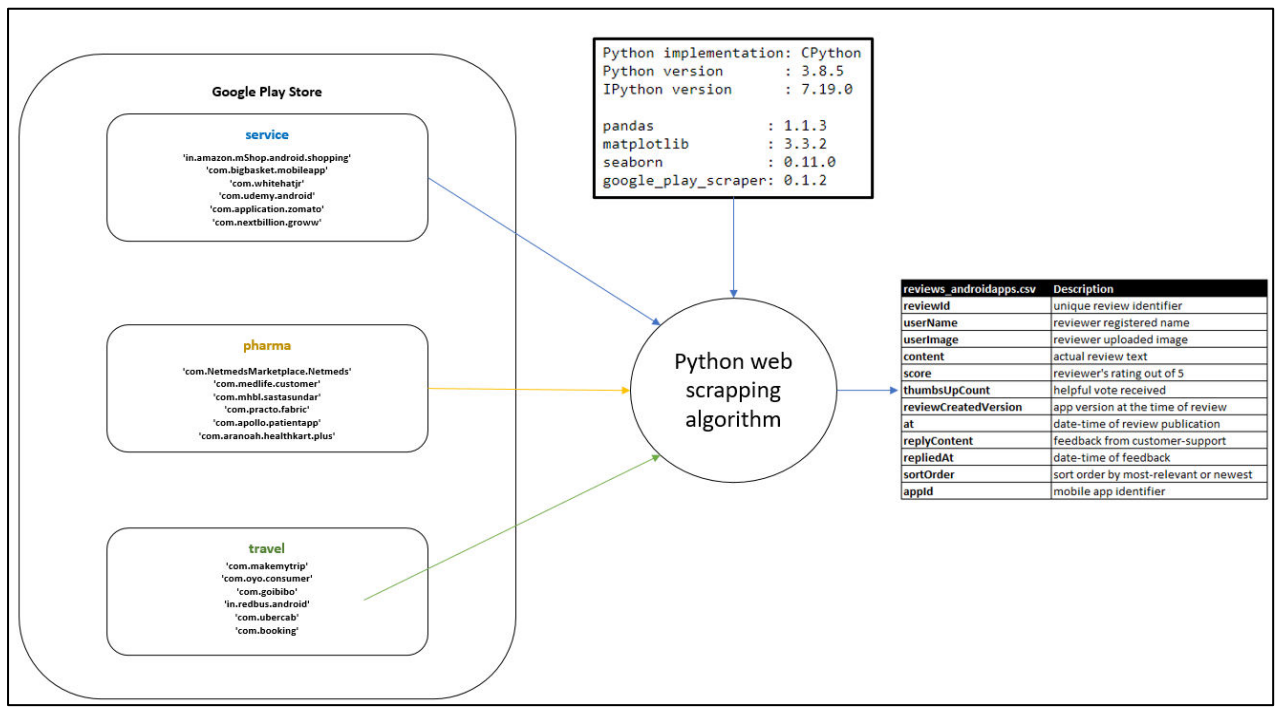

Figure 13. App-Review Web Scrapping Algorithm

For proper representation of the dataset, 1200 review comment comprise of 200 review comments with a rating scale of 3 while 100 review comments each of rating scale 1,2,4, and 5, each for a relevant and newest set. The metadata table 1 represents the datatypes that we collected with their description for further enrichment.

Table 1. Metadata of Scrapped Comma Separated Value (CSV) File

| reviews_androidapps.csv | Description |
|---|---|
| reviewId | unique review identifier |
| userName | reviewer registered name |
| userImage | reviewer uploaded image |
| content | actual review text |
| score | reviewer's rating out of 5 |
| thumbsUpCount | helpful vote received |
| reviewCreatedVersion | app version at the time of review |
| at | date-time of review publication |
| replyContent | feedback from customer-support |
| repliedAt | date-time of feedback |
| sortOrder | sort order by most-relevant or newest |
| appId | mobile app identifier |

**4.2 Enrichment Phase**

The data enrichment phase mainly comprised of sensitive data removal followed by addition of derived attributes. From the web data file, we retained content, score, thumbsUpCount, and appId attributes. Content attribute was used as an input to sentimentr package in r to generate commentSentiment attribute. We chose sentimentr package over plain lexical approach of using bing/nrc/afinn because we needed a sentence-level sentiment score as opposed to word-level score. Moreover, sentimentr takes in account of valence-shifters in a sentence that assigns a context-based sentiment by affecting the polarized words.

In our study we have used sentiment_by() function that calculates average sentiment after processing sentence level sentiment score for each review text. A -ve value denotes negative sentiment while a +ve value denotes positive sentiment. Our dataset had an average sentiment scores ranging from -1.91 to +2.07 spread over 21600 review text.

To get a toxicity score, we reuse the content attribute and pass it through the perspective application programming interface. Each review takes around 5 seconds to process and gives a value between 0 and 1. The fractional value denotes the probability of finding the content text as toxic. For example, 0.22 toxicity score means that there is a probability of 22 per 100 people finding the same text as toxic.

```r
## Step1: Call sentimentr and relevant data enrichment library
library(sentimentr)
library(tidyr)
library(tidyverse)

## Step2: Reading appstorereview data into dataframe with numeric id
df=read.csv("reviews_androidapps.csv",stringsAsFactors = FALSE)
df$id=1:dim(df)[1]
df_new=df %>%
  select("content", "score", "thumbsUpCount", "id")

## Step3: Splitting into dataframe df1 to df18 to MAP data
## for sentence-level aggregated sentiment scores
df1 <- df_new[1:1200,]
df1$sentiment <- sentiment_by(df1$content, by = NULL)$ave_sentiment
.
.
.
df18 <- df_new[20401:21600,]
df18$sentiment <- sentiment_by(df18$content, by = NULL)$ave_sentiment

## Step4: Appending df1 to df18 to REDUCE the final dataframe
df_full <- do.call("rbind", list(df1, df2, df3, df4, df5, df6, df7, df8,
df9, df10, df11, df12, df13, df14, df15, df16, df17, df18 ))

## Step5: Exporting final dataframe into external csv file
write.csv(df_full,"reviews_with_sentimentscore.csv", row.names = TRUE)
```

Figure 14: Comment Sentiment Scoring Algorithm in R

To get an optimized output from 21K textual content, a map-reduce framework was setup in R language as it supports both sentiment_by as well as perspective api. The entire data was split into 18 dataframes based on appID and then each of them were processed to add two new attributes namely commentSentiment and commentToxicity column. Streaming function were used to process toxicity score that handled the rate-limiting API calling constraint in perspective.

Sentiment score processing took approximately 6 hours to process and Toxicity score processing took approximately 30 hours to process. Finally, the 18 datasets were appended using rbind function.

```r
## Step1: Call API and relevant data enrichment library
library(peRspective)
library(tidyr)
library(tidyverse)
Sys.setenv(perspective_api_key = "AIzaSyCpPJmwyj_OuWgjbDu7-vPk_Apf6adI4eY")
## Step2: Reading appstorereview data into dataframe with alphanumeric id
df=read.csv("reviews_androidapps.csv",stringsAsFactors = FALSE)
df$id=1:dim(df)[1]
prefix <- "tox"
df$newid=paste(prefix, df$id, sep="")
df_new=df %>%
  select("content", "score", "thumbsUpCount", "appId", "newid")
## Step3: Splitting into Dataframe df1 to df18 to MAP API stream for Toxicity Scores
df1 <- df_new[1:1200,]
df1$commentToxicity <- (df1 %>%
                  prsp_stream(text = content,
                              text_id = newid,
                              score_model = c("TOXICITY"),safe_output = T,verbose = T))$TOXICITY
.
.
.
df18 <- df_new[20401:21600,]
df18$toxicity <- (df18 %>%
                  prsp_stream(text = content,
                              text_id = newid,
                              score_model = c("TOXICITY"),safe_output = T,verbose = T))$TOXICITY
## Step4: Appending df1 to df18 to REDUCE the final dataframe
df_full <- do.call("rbind", list(df1, df2, df3, df4, df5, df6, df7, df8,
df9, df10, df11, df12, df13, df14, df15, df16, df17, df18 ))
## Step5: Exporting final dataframe into external csv file
write.csv(df_full,"reviews_with_toxicityscore.csv", row.names = TRUE)
```

Figure 15: Comment Toxicity Scoring Algorithm in R

Post-processing data sanity and cleaning were done to remove redundant columns from the dataset along with 484 row removal which had NA toxicity score due to foreign language and illegible characters.

To ensure balanced dataset after enrichment, we conducted boxplot on the entire data as depicted in figure where serviceCluster 1,2, and 3 represents subscription-based-service, pharma, and travel respectively. The boxplot uniformity w.r.t ratingScore confirms the preservation of balanced-ness post processing of the dataset, as depicted in figure 16.

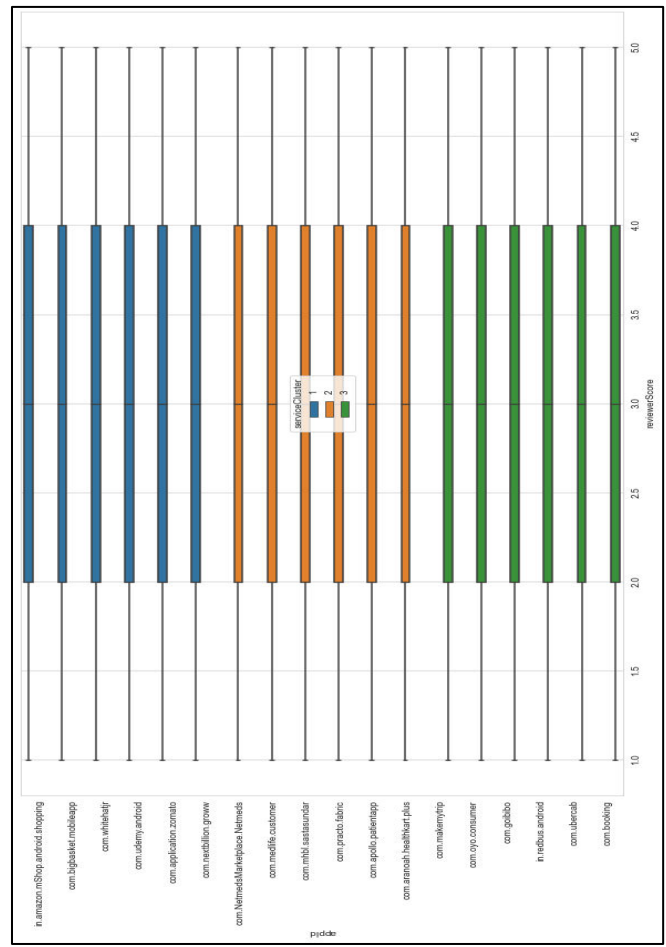

Figure 16: Sector-Wise Segregated Boxplot for various apps w.r.t reviewer-score.

**4.3 Evaluation Phase**

There are two basic functions that allows us to run path analysis in Lavaan: the 'sem' and the 'lavaan' functions (Thakkar, 2020). Our study specifies a path model involving only manifest/measured variables and estimates model parameters using the 'lavaan' function. The # (hash/pound) sign is used for comments and are not read by the program provided in appendix. Following steps are followed to generate the path model using lavaan package.

Pre-requisite: R packages lavaan, tidySEM, semplot, ggplot2, and dplyr needs to be installed from CRAN server with all dependencies.

Step 1: We create a dataframe named processdata by reading our comma separated value file from external folder.

Step 2: We use the command str to view the various datatypes in the dataframe processdata

Step 3: In the variable named model we define our path analysis definitions that has the regression and covariance operation

Step 4: We pass the model definition using processdata dataframe into the lavaan function and store the generated model in fit variable

Step 5: We use summary function to generate the model fit summary and graph_sem function to visualize the path model generated.

When specifying predictive relationships in the model, we use the tilde sign ('~'), which separates the dependent variable in each equation from its predictors. Predictors are separated in each equation by '+' sign. In our model, we also allow the residuals to correlate and is done by a double tilde ('~~') sign.

```
model<-'
#equation where commentSentiment is predicted by serviceCluster & reviewerScore and thumbsUpCount
commentSentiment~reviewerScore+thumbsUpCount+serviceCluster
#equation where serviceRelativeRating is predicted by commentSentiment and commentToxicity
serviceRelativeRating~commentToxicity+commentSentiment+reviewerScore
#equation where commentToxicity is predicted by reviewerScore and thumbsUpCount
commentToxicity~thumbsUpCount+reviewerScore
#estimtating the variances of the exogenous variables (serviceCluster, reviewerScore,thumbsUpCount)
reviewerScore~~reviewerScore
thumbsUpCount~~thumbsUpCount
serviceCluster~~serviceCluster
#estimtating the covariances of the exogenous variables (serviceCluster, reviewerScore,thumbsUpCount)
reviewerScore~~thumbsUpCount+serviceCluster
thumbsUpCount~~serviceCluster
#estimating the residual variances for endogenous variables (commentSentiment, commentToxicity, serviceRelativeRating)
commentSentiment~~commentSentiment
commentToxicity~~commentToxicity
serviceRelativeRating~~serviceRelativeRating
#estimating the covariance of residuals for commentSentiment and commentToxicity
commentSentiment~~commentToxicity'
```

Figure 17: Definition of path analysis model variable

In the path analysis model, we deal with only measured variables. As per the model, there are two kinds of measured variable namely endogenous and exogenous variable. The variables from which all the arrows are always pointing out are the exogenous ones while the other which have incoming arrows are the endogenous ones. The fit summary has been discussed in the model interpretation section.

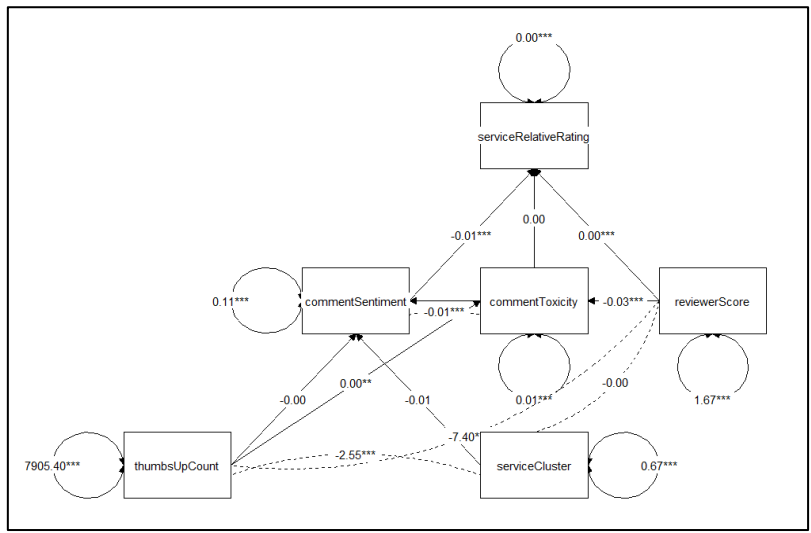

Figure 18: Graph-SEM representation of fit model

The measurement model is essentially a linear regression model where the main predictor, the factor, is latent or hidden or unobserved. Hence, the measurement model is also called the factor analysis model. There are three main differences between the measurement model and linear regression:

- Measurement model outcomes are not observations but constructs.
- Measurement model is a multivariate model. There are as many outcomes per subject as there are items. In a linear regression, there is only one outcome per subject.
- The factor, η ("eta"), is latent whereas in a linear regression the predictors are observed or measured.

In lavaan package, we use the symbol =~ to denote a measurement model. On the left side of the symbol we have the latent indicator while on the right side of the symbol we have the predictors as in regression model separated by the + sign, as shown in the figure.

```
measurementmodel<-'
#measurement model
sectorEffect =~ reviewerScore+thumbsUpCount+serviceCluster
productSentiment =~ commentSentiment+commentToxicity+serviceRelativeRating'
```

Figure 19: Definition of measurement model variable

Similar to path analysis model, we carry on the following steps for measurement model:

Pre-requisite: R packages lavaan, tidySEM, semplot, ggplot2, and dplyr needs to be installed from CRAN server with all dependencies.

Step 1: We create a dataframe named processdata by reading our comma separated value file from external folder.

Step 2: We use the command str to view the various datatypes in the dataframe processdata

Step 3: In the variable named measurementmodel we define our measurement model definitions using ~= operation

Step 4: We pass the model definition using processdata dataframe into the sem function and store the generated model in fitmm variable

Step 5: We use summary function to generate the model fitmm summary and semPaths function to visualize the measurement model generated in figure 20.

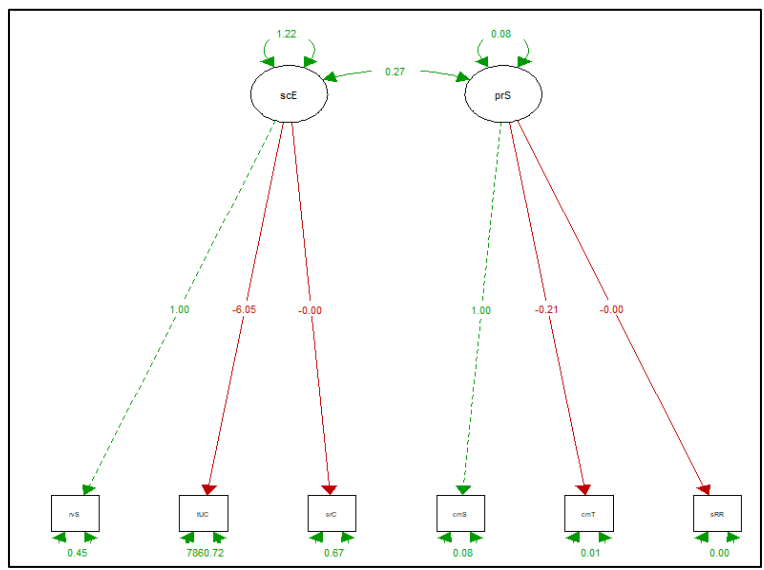

Figure 20: semPaths representation of fitmm model

Correlation heatmap in Figure 21 is also generated to get an idea of which measured variables are correlated. In the measurement model the oval shapes are the latent indicators while the rectangle represents the measured variables.

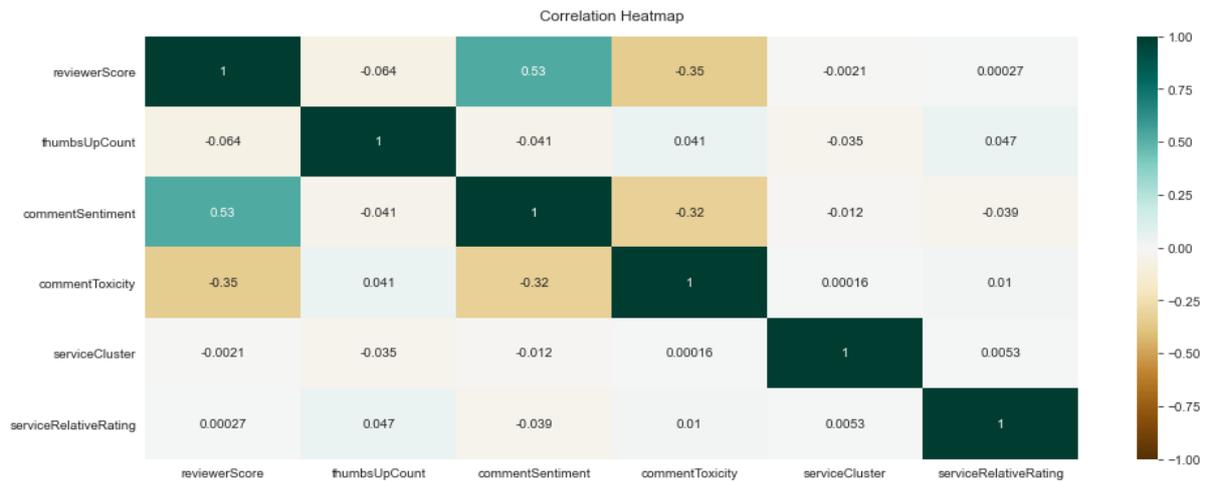

Figure 21: Correlation Heatmap

Total effect modelling or mediation modelling is a subset of SEM where we determine any mediating effects due to measurable variable. Mediation are derived as indirect effect. When added with the direct effect, it gives us the total effect model.

```
modelTE<-'
#labeling path from reviewerScore to commentSentiment
commentSentiment~a*reviewerScore+thumbsUpCount+serviceCluster
#labeling path from commentSentiment to serviceRelativeRating. Adding labeled path from reviewerScore to serviceRelativeRating
serviceRelativeRating~e*commentToxicity+b*commentSentiment+c*reviewerScore
#predicting commentToxicity and labeling path from reviewerScore
commentToxicity~thumbsUpCount+d*reviewerScore
#estimtating the variances and covariances of the exogenous
#variables (serviceCluster, reviewerScore,thumbsUpCount)
reviewerScore~~reviewerScore
thumbsUpCount~~thumbsUpCount
serviceCluster~~serviceCluster
reviewerScore~~thumbsUpCount+serviceCluster
thumbsUpCount~~serviceCluster
#estimating the variances of residuals for endogenous variables
#(commentSentiment, commentToxicity, serviceRelativeRating)
commentSentiment~~commentSentiment
commentToxicity~~commentToxicity
serviceRelativeRating~~serviceRelativeRating
#estimating the covariance of residuals for commentSentiment and commentToxicity
commentSentiment~~commentToxicity
#calculating specific indirect effect of reviewerScore on serviceRelativeRating via commentSentiment
SIE1:=a*b
#calculating specific indirect effect of reviewerScore on serviceRelativeRating via commentToxicity
SIE2:=d*e
#calculating total indirect effect of reviewerScore on serviceRelativeRating via mediators
TIE:=SIE1+SIE2
#calculating total effect of reviewerScore on serviceRelativeRating
TE:=TIE+c'
```

Figure 22: Definition of Total Effect model variable

In lavaan, mediation effects are shown by ": =" symbol. We use the following steps to determine the total effect model.

Pre-requisite: R packages lavaan, tidySEM, semplot, ggplot2, and dplyr needs to be installed from CRAN server with all dependencies.

Step 1: We create a dataframe named processdata by reading our comma separated value file from external folder.

Step 2: We use the command str to view the various datatypes in the dataframe processdata

Step 3: In the variable named modelTE we define our direct and indirect pathway definitions that has the := operator.

Step 4: We pass the model definition using processdata dataframe into the lavaan function and store the generated model in fitTE variable

Step 5: We use summary function to generate the model fit summary using naïve bootstrap for standard errors.

```
Indirect and total effects of reviewScore on serviceRelativeRating:

First specific indirect effect:

SIE1=path a*path b

Second specific indirect effect:

SIE2=path d*path e

Total indirect effect:

TIE=SIE1+SIE2

Total effect of reviewScore on serviceRelativeRating

TE=TIE + path c
```

Figure 23: Parameters of Total Effect model

We generate a Full SEM model to study the interaction between latent and measured variables. We used semPlots to visualize the interactions as depicted in Figure 25.

```
semFullModel<-'
#measurement model with marker variable identification
sectorEffect =~ 1*reviewerScore+thumbsUpCount+serviceCluster
productSentiment =~ 1*commentSentiment+commentToxicity+serviceRelativeRating
#regression model
productSentiment~sectorEffect'
```

Figure 24: Definition of Full SEM model

Pre-requisite: R packages lavaan, tidySEM, semplot, ggplot2, and dplyr needs to be installed from CRAN server with all dependencies.

Step 1: We create a dataframe named processdata by reading our comma separated value file from external folder.

Step 2: We use the command str to view the various datatypes in the dataframe processdata

Step 3: In the variable named semFullmodel we define our full SEM model using ~= and ~ operations, along with marker variable identification using *.

Step 4: We pass the model definition using processdata dataframe into the lavaan function and store the generated model in fitfull variable

Step 5: We use summary function to generate the model fit summary and semPaths function to visualize the generated model.

Figure 25: semPaths representation of semFullModel

## 5. Discussion

To effectively answer the research questions, we need to interpret the output of the models by lavaan package in R. We have walkthrough an infographic based on our path analysis model summary to understand the major sections in lavaan model fit summary.

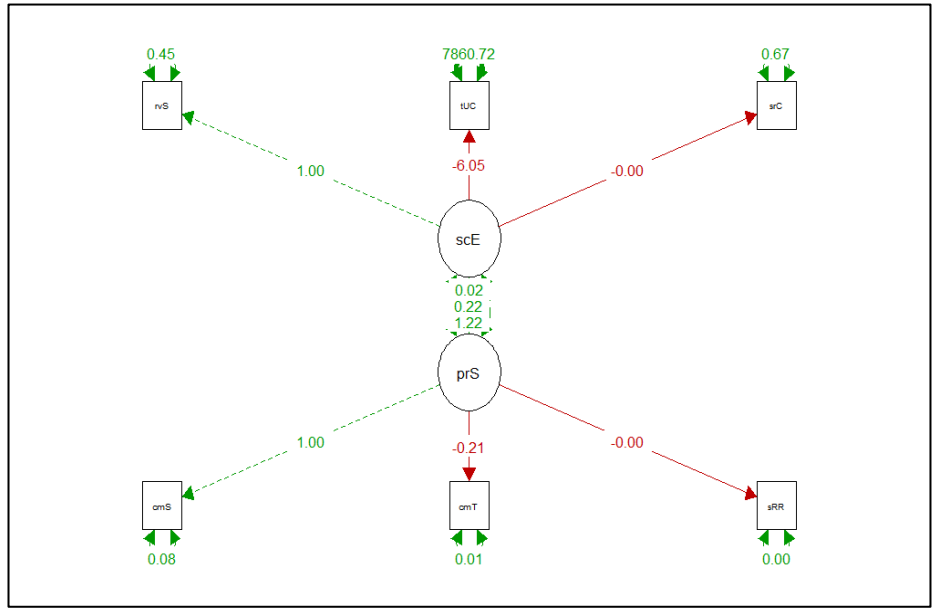

Figure 26: Section One of lavaan fit summary

The three sections are common to all the lavaan model outputs when the measure parameter is set to TRUE.

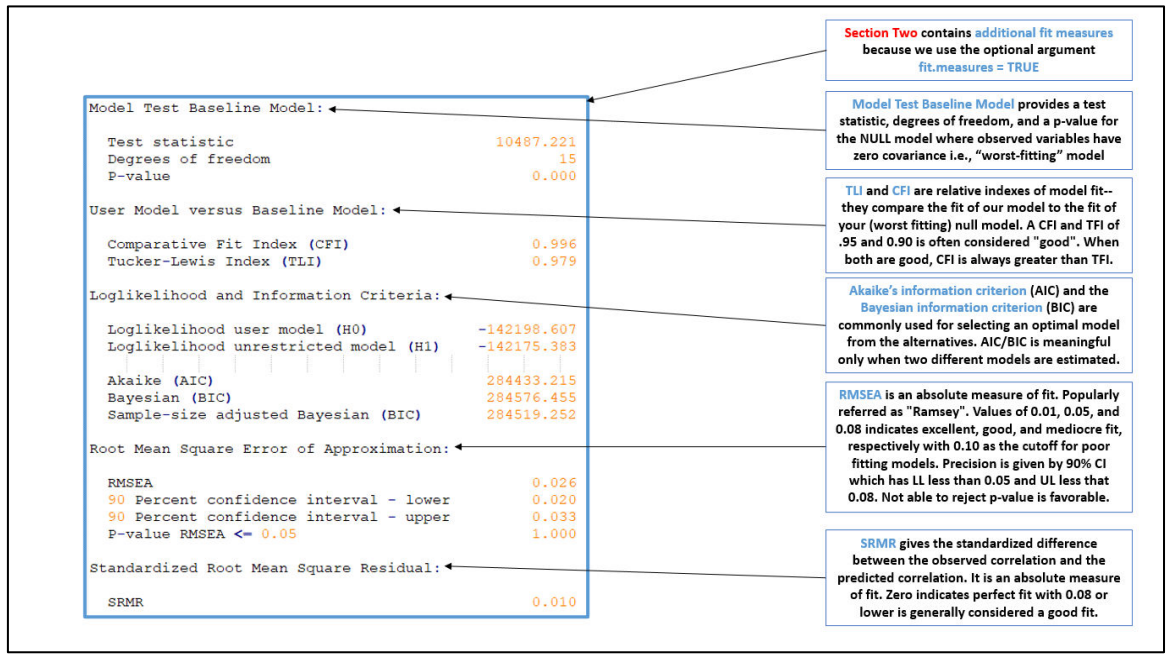

Figure 27: Section Two of lavaan fit summary

Although, the final section varies according to the parameters chosen to define the model, the nomenclature and way of interpreting is same as presented in the given infographics.

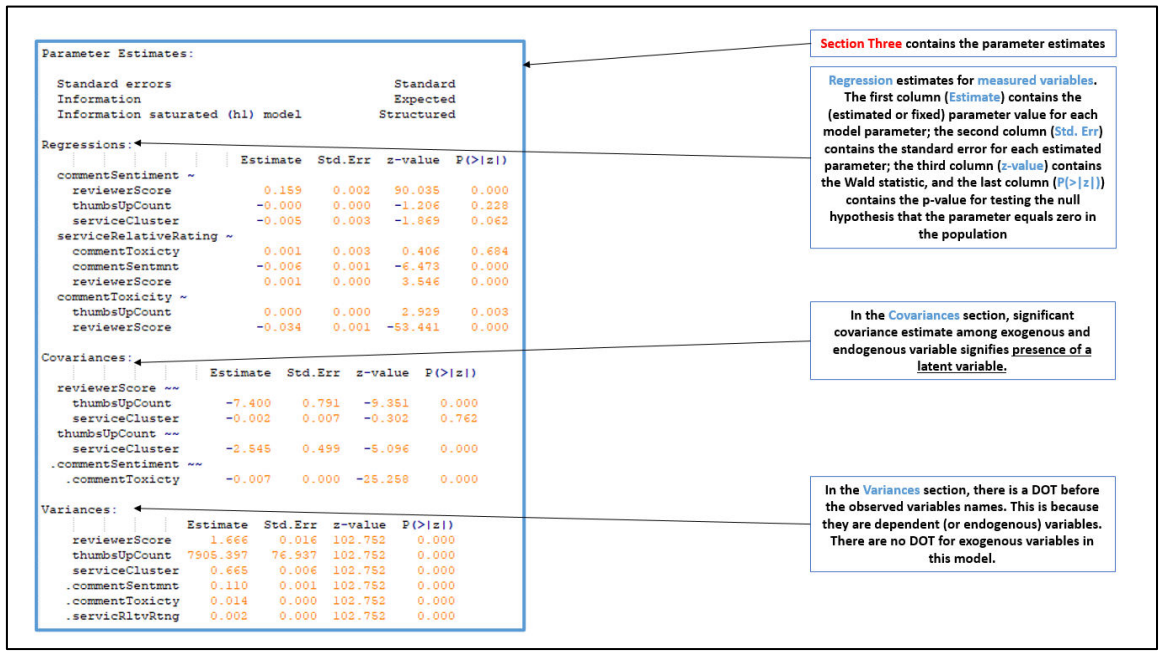

Figure 28: Section Three of lavaan fit summary

Before discussing our result, it is to be noted that our aim is not to develop a perfect model of variable interaction but to understand effect of the such an interaction. Hence, our study does not emphasize on a reliability test as we are not comparing various models to hunt down the perfect model. Rather, we concentrate on the statistical significance and fit summary of the models to check whether the proposed models are a better fit than the baseline null model. In order to accomplish this, we have summarized the necessary fit indices in figure 29 and table 2.

Table 2. Model Fit Summary

| Model | $\chi^2$ | df | CFI | TLI | RMSEA* | SRMR |
|---|---|---|---|---|---|---|
| Path Analysis (fit) | 46.449 | 3.000 | 0.996 | 0.979 | 0.026 | 0.010 |
| Measurement (fitmm) | 116.117 | 8.000 | 0.990 | 0.981 | 0.025 | 0.014 |
| Total Effect (fitTE) | 46.449 | 3.000 | 0.996 | 0.979 | 0.026 | 0.010 |
| Full SEM (fitFull) | 116.117 | 8.000 | 0.990 | 0.981 | 0.025 | 0.014 |

\* P-value RMSEA <= 0.05 is 1.000
\*\* ($\chi^2$)/df ignored for large dataset with p-value = 0.000

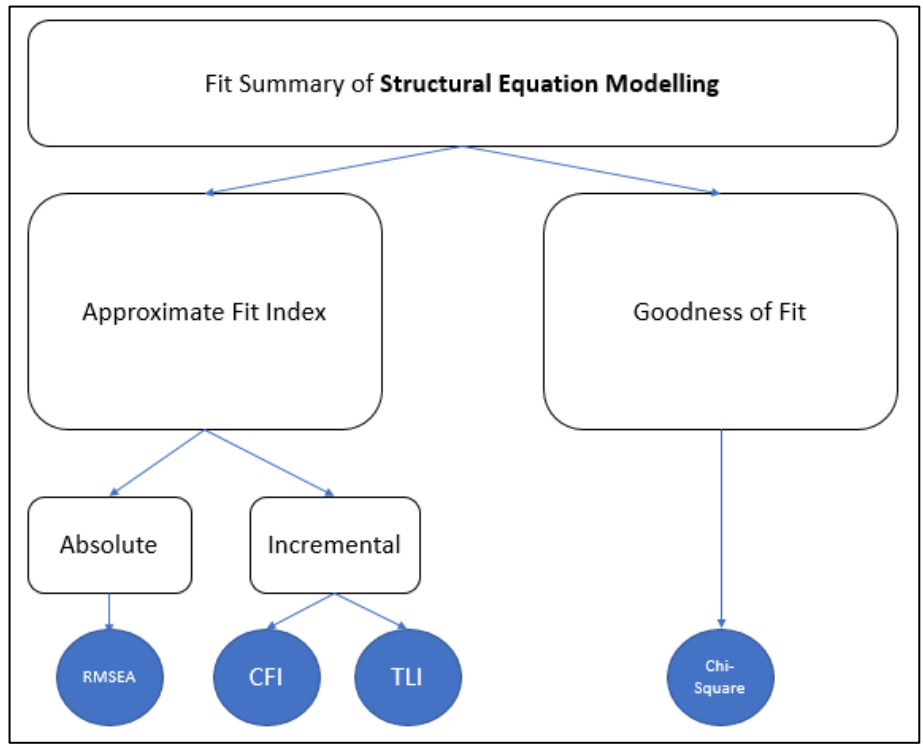

Figure 29: Model Fit Indices

For models with around 75 to 200 cases, the chi-square test is a reasonable measure of fit. In our study, with more cases (~1200), the chi-square is statistically significant. Hence, we mostly rely on degree of freedom (df) with non-zero value as a indication of non-saturated model rather than model accuracy. We also ignore chi-square to df ratio value and prefer Tucker-Lewis Index (TLI), Comparative Fit Index (CFI), Root mean square approximation (RMSEA), and standardized root mean square residuals (SRMR) for determining the model fit summary. In our infographics we have provided the acceptable range of TLI, CFI, and RMSEA. SRMR value less than 0.08 is considered a good fit (Hu & Bentler, 1999). As per the table 2 results, we can infer that the all the four models have a good fit compared to the baseline null model. Moreover, we see that Total effect model (fitTE) is built upon the path model (fit) while Full SEM model (fitFull) is built over the measurement model (fitmm).

**5.1 Discussion of Result**

In our study, the lavaan model summary outputs sections with regression and structural parameters. These parameters along with p-value can be used for model discussion and its significance.

```
Path Analysis Model

    Regressions:

    commentSentiment ~ 0.159*reviewerScore
    serviceRelativeRating ~ -0.006*commentSentmnt + 0.001*reviewerScore
    commentToxicity ~ -0.034*reviewerScore

Measurement Model:

    Latent Variables:

    sectorEffect =~ 1*reviewerScore -6.051*thumbsUpCount
    productSentiment =~ 1*commentSentmnt -0.212*commentToxicty -0.004*servicRltvRtng

Total Effect Model:

    SIE1 :=                a*b        (significant)
    SIE2 :=                d*e        (not-significant)
    TIE  :=          SIE1+SIE2        (significant)
    TE   :=              TIE+c        (not-significant)

Full SEM Model:

    Latent Variables:

    sectorEffect =~ 1*reviewerScore -6.051*thumbsUpCount
    productSentiment =~ 1*commentSentmnt -0.212*commentToxicty -0.004*servicRltvRtng

    Regressions:

    productSentiment ~ 0.217*sectorEffect
```

Figure 30: Model Implications (p-value < 0.005)

*Path Analysis model:*

With one unit increase in reviewer score we can have 0.159 increase in comment sentiment. Keeping reviewer score constant, with one-unit change in comment sentiment the service relative rating decrease by 0.006. With one unit increase in reviewer score we experience a drop of comment toxicity by 0.034.

*Measurement model:*

The latent indicator named sectorEffect decrease in loading by 6.051 for every unit increase in thumbsUpCount, keeping the reviewer score unchanged. The latent indicator named productSentiment decrease in loading by 0.212 for every unit increase in commentToxicity, keeping serviceRelativeRating and comment Sentiment constant.

*Total Effect model:*

Comment Sentiment significantly mediates reviewer score towards serviceRelativeRating. (path a * path b). Comment Toxicity have no significant mediation on reviewer score towards serviceRelativeRating. (path d * path e). There is a significant indirect effect of reviewer score on serviceRelativeRating but no significant direct effect.

*Full SEM model:*

Implications in measurement model remains unchanged. With a unit increase in sectorEffect indicator, there occurs a 0.217 increase in productSentiment indicator.

## 5.2 Implication

Based on the model discussions, we can implicate on the previously asked research questions.

*RQ1: Does the reviewer's score influences review content sentiment and toxicity?*

Increase in reviewer's score correlates to a positive content sentiment and a lowering of comment toxicity. This agrees to the AD model of consumer behaviour.

*RQ2: For a given review sentiment, does review toxicity influences overall product sentiment?*

With increase in review toxicity there is a decrease in overall product sentiment keeping the review sentiment score unchanged. This agrees with MPAA model of consumer behaviour.

*RQ3: Does comment toxicity acts as a significant mediator between reviewer's score and sector-wise relative rating?*

Comment toxicity has a non-significant mediation effect between reviewer's score and sector-wise review rating. This is because of the psychological distance a consumer perceives while availing various app-services of different sector. For example, in real emergencies, a consumer would anyhow use a pharma app irrespective of reviewer's rating.

*RQ4: Does overall product sentiment gets influenced by the sector of study?*

Overall product sentiment gets positively influenced by sector effect. This results correlates to the hypothetical aspect of construal level theory where the psychological distance drastically changes from a hypothetical situation to real-life experience among the intended consumer.

**5.3 Contribution**

Our study contributes along three key dimensions:

*Data Dimension*

Web data serves as an unbiased repository for understanding real-world dynamics. Data cleaning is an important aspect of using alternate data. Coupled with user sentiment, comment toxicity can serve as a quantitatively indicator the overall sentiment about a product.

*Process Dimension*

Bibliometric analysis serves as an objective method to conduct systematic literature review. Sector-wise analysis helps in identifying latent factors that influences overall product sentiment.

*Model Dimension*

Multivariate analysis helps in understanding complex dynamics of real-world interactions. Approximate Fit Index helps in identifying the Model better than goodness of fit Index. Full SEM model provides better insights compared to partial SEM models.

**5.4 Limitation and Future Scope**

Our study represents the tip of a massive iceberg of alternative data in form of product reviews and user generated comments that needs to be studied for social network analysis and socio-technical dynamics to enhance our perspective of both product and customer.

At each phase of our study, we made certain assumptions that limited our area of study that has been summarized below:

- Our data collection time-window was hard-bound. Present data changes were not considered.

- Sentiment scoring and Toxicity scoring were generated using application programming interface. Any unintentional bias inbuilt in these API tools would not be detected.
- We have used alternative dataset for only application review data rather than actual product review for the analysis. Only English based review was part of dataset.
- We have used only one base-app for collecting sector-wise data. Other app-aggregators were not used.
- We used covariance-based SEM methods for model evaluation. Partial least square method was not done.
- The relevant and newest user parameter has not been taken into context for the present study. A relevant user is such a user who has a commented on other apps frequently with sufficient helpful-vote counts. Such a study can generate further insight on a hypothetical construal level.

Beyond our study, we propose three major scopes for future NLP based SEM framework:

A review comment can be non-toxic but still repulsive. These involve profanity, sarcasm, attack-to-author, or plain irrelevance. In our study we only scored on the basis of toxicity. Future work can consider all the above-mentioned aspects for a holistic framework of interaction.

Many languages have gender-neutral terms, that when translated in English can induce severe unintentional bias. Future work can involve the preservation of gender-neutral terms while translation of review content that can influence toxicity or sentiment scoring.

To increase model accuracy and validation, future work can be approached using splitting up of data into training and test dataset. Such data-splits can help in inculcating Lasso and Ridge styled regularization for enhancing the robustness of the framework on new alternative data.

Proceedings of the 2017 ACM on Web Science Conference - WebSci '17. doi:10.1145/3091478.3091487

Chen, H., & Li, S. (2018). Measuring the Psychological Distance between an Organization and Its Members—The Construction and Validation of a New Scale. Frontiers in Psychology, 8. https://doi.org/10.3389/fpsyg.2017.02296

Cheng, J., Bernstein, M., Danescu-Niculescu-Mizil, C., & Leskovec, J. (2017). Anyone Can Become a Troll: Causes of Trolling Behavior in Online Discussions. CSCW : proceedings of the Conference on Computer-Supported Cooperative Work. Conference on Computer-Supported Cooperative Work, 2017, 1217–1230. https://doi.org/10.1145/2998181.2998213

DeSmet, A., Bastiaensens, S., Van Cleemput, K., Poels, K., Vandebosch, H., Cardon, G., & De Bourdeaudhuij, I. (2016). Deciding whether to look after them, to like it, or leave it: A multidimensional analysis of predictors of positive and negative bystander behavior in cyberbullying among adolescents. Computers in Human Behavior, 57, 398–415. https://doi.org/10.1016/j.chb.2015.12.051

Djuric, N., Zhou, J., Morris, R., Grbovic, M., Radosavljevic, V., & Bhamidipati, N. (2015, May). Hate speech detection with comment embeddings. In Proceedings of the 24th international conference on world wide web (pp.29-30). ACM.

Garfield, E. (2004). Historiographic Mapping of Knowledge Domains Literature. Journal of Information Science, 30(2), 119–145. doi:10.1177/0165551504042802

Greenberg, A. "Now anyone can deploy Google's troll-fighting AI", Wired, Feb. 23, 2017,

**Appendix**

Following list consist of code and data used in this paper, **available on request**:

| File | Type | Tool | Description |
| --- | --- | --- | --- |
| python_review_scrapper_algorithm.py | Code | Python | Algorithm for scrapping on app-reviews from google-playstore |
| comment_sentiment_scoring_code.r | Code | R | Algorithm for Sentiment Scoring on review content |
| comment_toxicity_scoring_code.r | Code | R | Algorithm for Toxicity Scoring on review content |
| python_review_sentox_data_analysis.pdf | Code | Python (Jupyter Notebook) | Data Analysis and visualization on Sector-wise review data |
| sem_model_summaries.r | Code | R | Algorithms for Structural Equation Modelling with model summary |
| reviews_androidapps.csv | Data | MS Excel (csv) | Raw Data scrapped from google-playstore on app-reviews |
| reviews_with_sentimentscore.csv | Data | MS Excel (csv) | Data with Sentiment Scores on review content |
| reviews_with_toxicityscore.csv | Data | MS Excel (csv) | Data with Toxicity Scores on review content |
| serviceRelativeRating.xlsx | Data | MS Excel (xlsx) | Sector-wise Relative Rating on Service, Pharma, and Travel |
| reviews_with_sentox_data.csv | Data | MS Excel (csv) | Combined data for sentiment and toxicity score |
| reviews_with_sentox_sectorwise.csv | Data | MS Excel (csv) | Cleaned data with sentiment score, toxicity score and sector-wise rating |